\documentclass[useAMs,usenatbib,a4paper]{mn2e}
\usepackage{savesym}
\usepackage{graphicx}
\expandafter\let\csname equation*\endcsname\relax
  \expandafter\let\csname endequation*\endcsname\relax 
\usepackage{subfig}
\usepackage{amsmath}
\usepackage{amssymb}
\usepackage{verbatim}
\usepackage{array}
\usepackage{times}
\usepackage[total={17.8cm,24.0cm},centering]{geometry}

\newcommand{\degree}{\ensuremath{^\circ}}

\newcommand{\be}{\begin{equation}}
\newcommand{\beq}{\begin{equation}}
\newcommand{\ee}{\end{equation}}
\newcommand{\eeq}{\end{equation}}
\newcommand{\eea}{\end{eqnarray}}
\newcommand{\bea}{\begin{eqnarray}}

\newcommand\bb[1] { \mbox{\boldmath{$#1$}} }
\newcommand\bcdot{\bb{\cdot}}

\def\gtsima{$\; \buildrel > \over \sim \;$}
\def\gtsim{\lower.5ex\hbox{\gtsima}}

\def\ltsima{$\; \buildrel < \over \sim \;$}
\def\ltsim{\lower.5ex\hbox{\ltsima}}

\title[The GSF instability in stellar interiors]{The Goldreich-Schubert-Fricke instability in 
stellar radiative zones}

\author[A. Caleo et al.]{
Andrea Caleo$^{1}$\thanks{E-mail: andrea.caleo@physics.ox.ac.uk}, Steven A. Balbus$^{1}$ and Emanuele Tognelli$^{2,3}$ \\
$^{1}$Oxford Astrophysics Department, Denys Wilkinson Building, Keble Road, Oxford, OX1 3RH, United Kingdom\\
$^2$Physics Department ``E. Fermi'', University of Pisa, Largo Bruno Pontecorvo 3, I-56127 Pisa, Italy \\
$^{3}$INFN, Section of Pisa, Largo Bruno Pontecorvo 3,  I-56127 Pisa, Italy \\
}

\begin{document}
\date{}

\pagerange{\pageref{firstpage}--\pageref{lastpage}} \pubyear{2015}

\maketitle

\label{firstpage}

\begin{abstract}
The Goldreich-Schubert-Fricke (GSF) instability is a rotational instability that is thought to contribute to the transfer of angular momentum in differentially rotating stars. It has been included in recent codes of stellar evolution in a diffusion-like approximation, under the assumption that the kinematic viscosity $\nu$ is unimportant for the development of the instability.  As noted previously by other authors, for most stellar applications this may not be a valid approximation. We discuss this issue in detail, solving the dispersion relation of the perturbed modes for realistic values of $\nu$ in the bulk of the radiative zone of the Sun and of three red giant stars at various evolutionary stages. We find that the GSF instability is triggered only in layers of extremely strong shear. In a simple case study, we also investigate the effect of a small deviation from axisymmetry or a small background magnetic field. We find that, like the viscosity, these have a stabilising effect. We conclude that this instability is probably far less efficient in transporting angular momentum than is often assumed, and may not even be present.
\end{abstract}


\section{Introduction} \label{sec:intro}
The Goldreich-Schubert-Fricke (GSF) instability (\citealt{GoldreichSchubert1967}, hereafter GS67, and \citealt{Fricke1968}) is an axisymmetric hydrodynamic instability that occurs in a shearing background when thermal diffusion from the displaced fluid elements counterbalances the stabilising effect of a positive entropy gradient.  Since the advent of helioseismology and asteroseismology and their recent observational achievements, it has gained attention in the recent literature as one of the effects that are thought to contribute to the transfer of angular momentum (AM) inside stars.

Stellar evolution codes (e.g. MESAstar, see \citealt{Paxton2013}) incorporate the GSF instability together with other hydrodynamic and magnetic processes, though not in the form of fundamental dynamics. These effects are modelled as a diffusive term in the equation of the evolution of the AM distribution. The combined effect of these processes is still unable to reproduce the correct AM flux for subgiants and early RGs (red giants) \citep{Cantiello2014}---at least in the current implementation. Codes are becoming more comprehensive, and other effects, including wave transport, may also soon be included. This approach holds promise, but the physics of the mechanisms that transport AM is often complicated.  Increasingly constraining observations from asteroseismology motivate us to revisit these mechanisms and to re-examine with critical attention the approximations used in their modelling.   In this paper in particular, we revisit the GSF instability.

The circulation induced by the development of the GSF instability is also thought to play a role in the chemical evolution of stars on the asymptotic giant branch (AGB), see e.g. \citet{Herwig2003}, \citet{Siess2004}.  These stars are known to be the main producers of elements heavier than iron, but theoretical models continue to struggle to reproduce the large spread in the heavy elements distribution observed among various AGB stars. \citet{Piersanti2013} attribute this phenomenon to rotational effects, and in particular to the mixing induced by the onset of the GSF instability at the top of the radiative zone.  Revisiting the physics of the GSF instability will have an impact on our understanding of the chemical evolution of evolved stars.

The manner in which the onset of the GSF instability is thought to cause a sort of zonal flow, which we will refer to as GSF circulation, is detailed in \citet{JamesKahn1970,JamesKahn1971} and is the basis of the diffusion-like approximation. The authors argue that the GSF instability is self-limiting, and provide an estimate for the AM flux in the fluid, after it has adjusted to suppress the growth of the unstable modes. This situation may be analogous to thermohaline convection (i.e. convection driven by destabilising compositional gradients) in low-mass red giants - numerical simulations in the low Prandtl number case have found that a linearly unstable fluid generates only small fingers,  which only have a modest effect in mixing the material in the star \citep{Denissenkov2011}.

The results by James \& Kahn were then taken up by \citet{EndalSofia1978} in the first evolutionary study of rotating stars with time-dependent redistribution of angular momentum.  Since then, the formula for the AM flux due to the GSF circulation has become a standard part of many evolutionary codes.  As of this writing, there are no simulations that either confirm or contradict the existence of this circulation.   Whether it truly exists and its effectiveness in transporting angular momentum in a star are largely unknown.

One of the approximations made by James and Kahn is that viscosity is ignored.    However, an important but subtle feature of the GSF instability is that it is can be suppressed by the presence of a very small kinematic viscosity $\nu$, even if the Prandtl number Pr = $\nu / \chi$, where $\chi$ is the thermal diffusion coefficient, is much smaller than unity \citep{Acheson1978, KnoblochSpruit1982, Menou2004}. In a recent paper, we have showed that it is likely that the GSF instability is suppressed in the upper radiative zone of the Sun, if the shear is comparable in value to that inferred from helioseismology \citep[hereafter CB16]{CaleoBalbus2016}. In this paper, we will explore in greater detail the conditions under which the GSF instability occurs in differentially rotating stars. We find that when realistic values of $\nu$ are considered, the instability is triggered only in regions of very strong shear, and we determine the minimum shear required to trigger the instability in the radiative zone of the Sun and of three RGs at various evolutionary stages. 

Recently, a similar study was conducted by \citet{Hirschi2010}. These authors have shown that the GSF instability would be suppressed by a turbulence-induced viscosity $\nu_{\text{turb}}$ in the radiative interior of a massive star. Here we consider instead a physical viscosity, i.e. the molecular viscosity and radiative viscosity. Our understanding of these processes is on a firmed footing than those underlying $\nu_{\text{turb}}$.

In a particularly simple case, we also investigate what happens when two of the assumptions behind the work by Goldreich and Schubert are relaxed. We consider a generic GSF-unstable environment, and then include (a) a small deviation from axisymmetry in the form of a finite azimuthal component $k_\phi$ of the wave vector, and (b) a small background magnetic field. 

This paper is organised as follows. Section 2 revisits the basics of the GSF instability. Section 3 presents a numerical study of the dispersion relation by GS67 with realistic values of the kinematic viscosity in a variety of stellar environments. Section 4 presents our investigation of the effect of a small deviation from axisymmetry or the presence of a small background magnetic field on an otherwise unstable environment. Section 5 summarizes our results.

\section{The Goldreich-Schubert-Fricke instability}   \label{sec:basicideas}
We make use both of standard spherical coordinates $(r, \phi, \theta)$ as well as cylindrical coordinates $(R, \phi, z)$. Throughout this paper, we consider a background angular velocity field which is azimuthally symmetric but otherwise arbitrary: $\Omega = \Omega (r, \theta)$ or $\Omega = \Omega(R, z)$.

In their seminal paper, Goldreich and Schubert performed a WKB analysis of modes with wavelength $\lambda \ll R$ allowing for finite thermal diffusion and kinematic viscosity in the medium. The perturbations depend on space and time as:
\begin{equation}
\exp[q \Omega t +\text{i}(k_R R + k_z z)] ,
\end{equation}
where $t$ is time, $q$ is a (complex valued) wave frequency, and $\bb k = (k_R, 0, k_z)$ is the wave vector. The system is then unstable if the dispersion relation has solutions with a positive real part of $q$.  For ease of reference, we reproduce here in a compact form the dispersion relation describing the evolution of the modes:
\begin{equation} \label{GS0}
q^3 + A(\bb k) q^2 + B(\bb k) q + C(\bb k) = 0 ,
\end{equation}
where
\begin{equation} \label{GS1}
A(\bb k) = \frac{k^2}{\Omega} \Big( 2\nu +  \frac{1}{\gamma} \chi \Big), 
\end{equation}
\begin{equation} \label{GS2}
	\begin{aligned}
B(\bb k) = & - \Big( \frac{k_z}{k} \Big)^2 \Big[ \frac{1}{\gamma \Omega^2 \rho} (\widetilde D P) (\widetilde D \sigma) +  \frac{2}{\Omega R} \widetilde D l \Big] + & \\
& + \frac{2}{\gamma} \Big( \frac{\chi  k^2}{\Omega} \Big)  \Big( \frac{\nu k^2}{\Omega} \Big) ,
	\end{aligned}
\end{equation}
\begin{equation} \label{GS3}
	\begin{aligned}
C(\bb k) = & - \Big( \frac{k_z}{k} \Big)^2 \Big( \frac{\nu k^2}{\Omega} \Big) \Big[ \frac{1}{\gamma \Omega^2 \rho}  (\widetilde D P) (\widetilde D \sigma) \Big] - & \\
	& - \Big( \frac{k_z}{k} \Big)^2 \Big( \frac{\chi  k^2}{\Omega} \Big) \Big[ \frac{2}{\gamma \Omega R}   \widetilde D l  \Big] + & \\
	& + \frac{1}{\gamma} \Big( \frac{\chi  k^2}{\Omega} \Big) \Big(\frac{\nu k^2}{\Omega} \Big)^2   .
	\end{aligned}
\end{equation}
In these equations, $\rho$ is the density of the fluid, $P$ its pressure, $T$ its temperature, $\nu$ the kinematic viscosity, $\sigma = \log{(P \rho^{-\gamma})}$ the entropy variable, $\gamma$ the adiabatic index, $\chi$ the heat conductivity, $l = \Omega R^2$ the angular momentum per unit mass, and $k^2 = k_R^2 + k_z^2$. We introduced the differential operator \citep{Balbus1995}:
\begin{equation}
\widetilde D = \frac{k_R}{k_z} \frac{\partial}{\partial z} - \frac{\partial}{\partial R} .
\end{equation}

Inspection of equation \eqref{GS0} immediately shows that there is at least one real positive solution for $q$ if $C(\bb k) < 0$ for any $\bb k$. A necessary stability criterion is therefore $C(\bb k) > 0$. Goldreich and Schubert focused on the small Prandtl number case: Pr = $\nu / \chi \rightarrow 0$. In this case, the stability condition reduces to $\widetilde D l < 0$, or in other words: 
\begin{equation}
\frac{k_R}{k_z} \frac{\partial l}{\partial z} - \frac{\partial l}{\partial R} < 0.
\end{equation}
Since the ratio $k_R / k_z$ can assume any value (either positive or negative), this condition can hold only if
\begin{equation} \label{GScond}
\frac{\partial l}{\partial R} > 0 \qquad \text{and} \qquad \frac{\partial \Omega}{\partial z} = 0 ,
\end{equation}
i.e.the angular momentum increasing outward and the angular velocity fixed on cylinders. This is the criterion of Goldreich and Schubert, and it provides a very strong constraint: any rotation pattern with $\partial_z \Omega \ne 0$ would be subject to an exponentially growing instability. A naive application of equation \eqref{GS0} with $\nu = 0$ to the upper radiative zone of the Sun gives a growth time-scale of less than 10 years (CB16). Models with $\Omega$ fixed at a given spherical distance from the centre, $\Omega = \Omega(r)$ (``shellular'' rotation, see \citealt{Meynet1997}) are often used in modern stellar codes; all these structures would be unstable unless the star rotates as a rigid body. As we have mentioned in the introduction, however, it was soon realised that the instability may be self-limiting and its consequences less dramatic.

In the radiative zone of the Sun Pr $= \nu/\chi \sim 10^{-6} - 10^{-5}$, so it might appear at first sight that one could neglect the viscosity when studying the onset of the instability.   However, as reported above and elsewhere, the ratio of the terms that multiply $\nu$ in the first line of the equation to those that multiply $\chi$ in the second line is large.   This number is about $10^5$ in the radiative zone of the Sun, so that the neglect of the viscous terms is not a good approximation.

\section{The GSF instability in various stellar models} \label{sec:numericalresults}
We discuss here the onset of the GSF instability in stellar environments, retaining the viscosity in equation \eqref{GS3}.  
For our background states, we have computed evolutionary tracks for the Sun and for a set of RGs.   We selected the models of a 1.3 M$_\odot$ star in three different evolutionary states on the RGB (red giant branch), corresponding to a subgiant, an early RG, and a more evolved RG, following the example by \citet{Belkacem2015b}.

\subsection{Stellar models} \label{subsec:stellarmodels}
We have generated a set of stellar models with the evolutionary code \texttt{PROSECCO} (see \citealt{Tognelli2012}). The main, rather standard, input physics adopted for the computations is detailed in \citet{Tognelli2015b} and references therein. The code generates spherically symmetric stellar models in hydrostatic equilibrium. Convection in super adiabatic regions is treated by the Mixing Length scheme \citep{MixingLength1958}. The models are computed adopting a solar-calibrated value of the mixing length parameter, namely $\alpha_\text{ML} = 1.76$. We adopted a mild overshooting parameter $\lambda_\text{ov}=0.2$ for $M\ge 1.2$ M$_\odot$. All the models are calculated for [Fe/H]=$+0.0$, which translates into an initial helium abundance $Y=0.274$ and a total metallicity $Z=0.013$.

\begin{figure}
	\centering
	\includegraphics[width=0.475\textwidth, clip=true, trim=0cm 0cm 0cm 0cm]{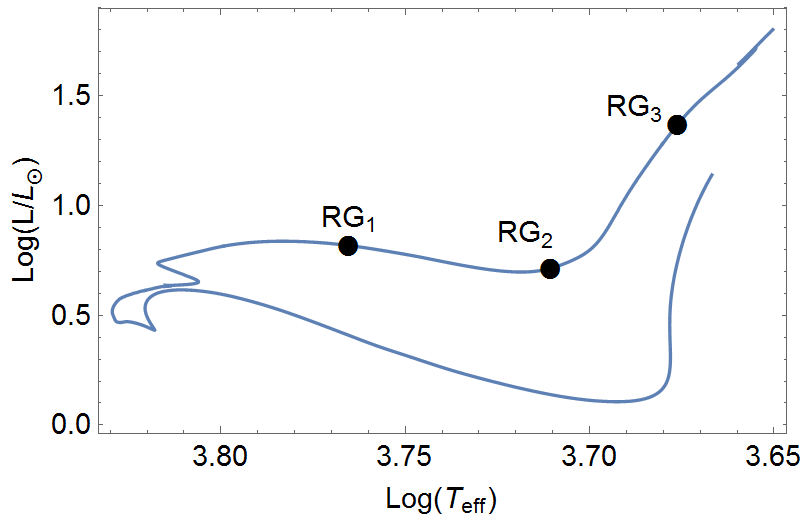}					
	\caption{\label{figHR}Evolutionary track of a 1.3 M$_\odot$ star in the HR diagram. The location of the three selected models is marked in the figure. $T_{\text{eff}}$ is expressed in K.}
\end{figure}

Figure \ref{figHR} shows the evolutionary track of the 1.3 M$_\odot$ star in the effective temperature ($T_{\rm eff}$) luminosity ($L$) plane, and the position of the three selected models in the HR diagram.  The main properties of our models are summarised in table \ref{tab:models}. The last column of the table shows the radius of the radiative region of the star as a fraction of the total radius. In each case, the generation of energy via nuclear fusion, be it in the core (main sequence) or in shell (RGs), occurs entirely in a zone of radiative transport.

\begin{table}
\centering
\begin{tabular}{ c  c  c  c c  c }
\hline \hline
 Model   &   $t$ (Myr)   & $R/R_\odot$ & $L/L_\odot$ & $T_{\text{eff}}$ (K) &  $R_{\text{rad}} / R$ \\ \hline
Sun &	 4570  & 1.00	&	1.00 &  5776	&  0.72 \\
RG1 & 3610 & 	2.52 &	6.56 &	5822 & 0.70	\\
RG2 & 3730	& 2.88& 5.18	& 5133 & 0.42	\\
RG3 & 3995 & 7.16	& 23.4	& 4744	& 0.08	\\ \hline
\end{tabular}
\caption{\label{tab:models}Main characteristics of the selected models: age, radius, luminosity, effective temperature, and location of the radiative-convective boundary.}
\end{table}

\subsection{Diffusive processes} \label{sec:diff}
Whether or not the GSF instability occurs depends on the microscopic diffusion coefficients of the stellar material.

The heat flux in the radiative region of a star is of course due to diffusive radiative transfer.  Adopting the same notation of GS67, the thermal conductivity for this process is:
\begin{equation} \label{chi}
\chi = \frac{16 (\gamma - 1) \bar m \sigma_{\text{SB}}  T^3}{3 k_\text{B} \kappa \rho^2},
\end{equation}
where $\sigma_{\text{SB}}$ is the Stephan-Boltzmann constant, $k_\text{B}$ is the Boltzmann constant, $\gamma$ is the adiabatic index, $\bar m$ is the average particle mass, and $\kappa$ is the opacity.

Two processes contribute to the viscosity in the star: the diffusion of particles and the diffusion of photons.  We follow the convention of referring to the first as ``molecular'' viscosity (though the particles are of course highly ionised atoms) and to the second as radiative viscosity. The molecular viscosity is given by \citep{Spitzer1962}:
\begin{equation} \label{numol}
\nu_{\text{dyn}} \cong 2.2 \times 10^{-15} \frac{T^{5/2}}{\log(\Lambda) \rho}  \text{ cm$^2$ s$^{-1}$},
\end{equation}
where $\log(\Lambda)$ is the Coulomb logarithm. Values for $\log(\Lambda)$ given the physical properties of the plasma are tabulated by \citet{Spitzer1962}. We adopt $\log(\Lambda) = 4$, an acceptable approximation throughout the radiative zone of the Sun and in the radiative zone of the RGs.

The radiative viscosity is given by (GS67; see also \citealt{Thomas1930}):
\begin{equation} \label{nurad}
\nu_{\text{rad}} = \frac{16 \sigma_{\text{SB}} T^4}{15 c^2 \kappa \rho},
\end{equation}
where $c$ is the speed of light. Inspection of equations \eqref{chi} and \eqref{nurad} shows that the ratio between $\nu_{\text{rad}}$ and $\chi$ is of order:
\begin{equation} \label{chiovernurad}
\frac{\nu_{\text{rad}}}{\chi} \sim \Big(\frac{c_\text{s}}{c}\Big)^2 ,
\end{equation}
where $c_\text{s}$ is the isothermal sound speed. The molecular viscosity
$\nu_{\text{dyn}}$ is usually more important than $\nu_{\text{rad}}$ in stars. A notable exception, which is relevant to this paper, is given by the core of RGs. Equation \eqref{chiovernurad} provides a lower limit for the Prandtl number, but in most cases it is not a good approximation for it.

\begin{figure}
	\centering
	\includegraphics[width=0.475\textwidth, clip=true, trim=0cm 0cm 0cm 0cm]{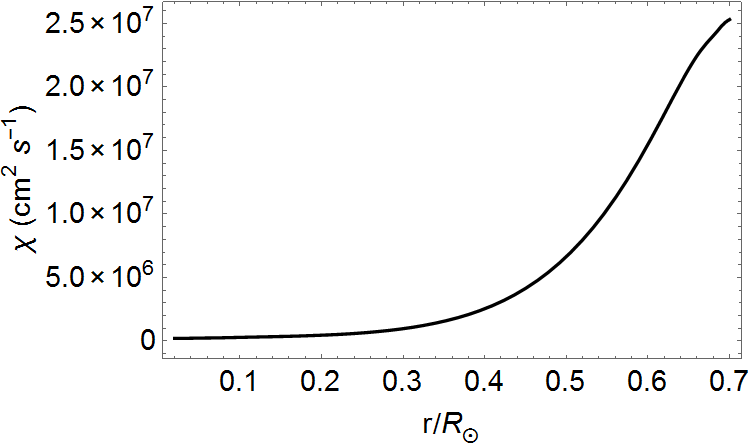}					
	\caption{\label{figChiSun}Thermal conductivity $\chi$ in the radiative zone of the Sun.}
\end{figure}

\begin{figure}
	\centering
	\includegraphics[width=0.475\textwidth, clip=true, trim=0cm 0cm 0cm 0cm]{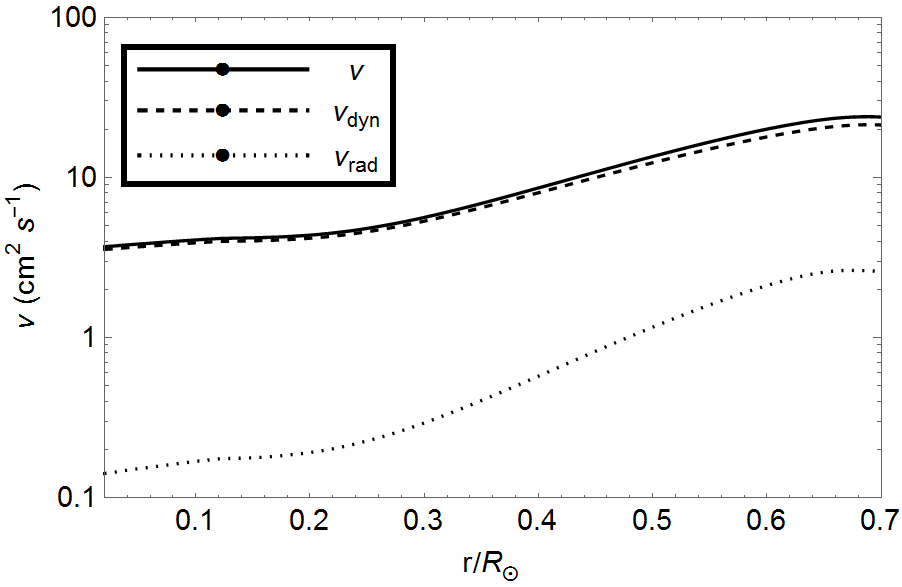}					
	\caption{\label{figNuSun}Kinematic viscosity $\nu$ in the radiative zone of the Sun.}
\end{figure}

\begin{figure}
	\centering
	\includegraphics[width=0.475\textwidth, clip=true, trim=0cm 0cm 0cm 0cm]{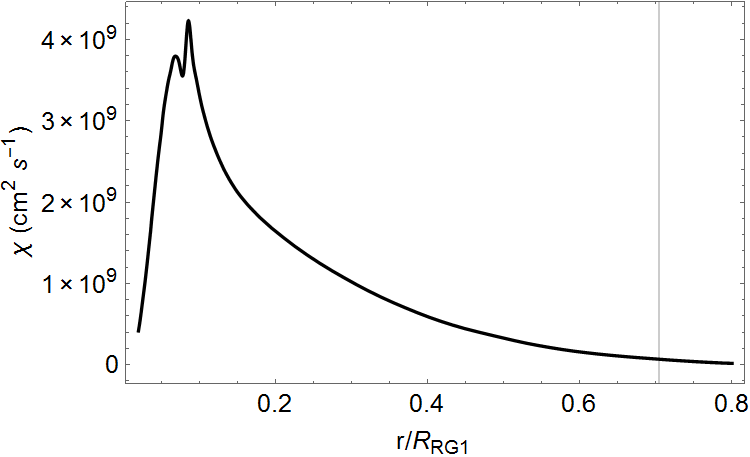}					
	\caption{\label{figChiRG}Thermal conductivity $\chi$ in the RG1 model. The vertical grey line shows the boundary between the radiative and convective zone.}
\end{figure}

\begin{figure}
	\centering
	\includegraphics[width=0.475\textwidth, clip=true, trim=0cm 0cm 0cm 0cm]{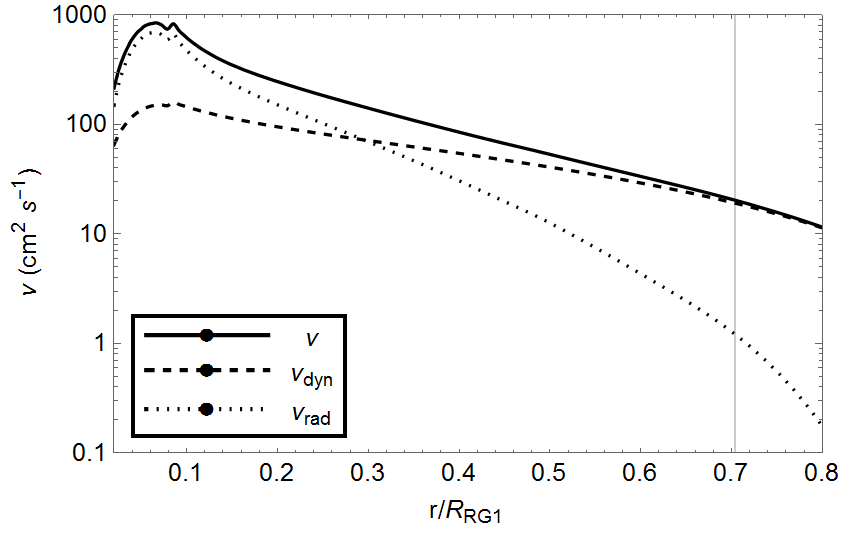}					
	\caption{\label{figNuRG}Kinematic viscosity $\nu$ in the interior of the RG1 model. The vertical grey line shows the boundary between the radiative and convective zone.}
\end{figure}

The thermal conductivity and kinematic viscosity in the Sun are shown in figures \ref{figChiSun} and \ref{figNuSun}; the same quantities are shown for the model RG1 in figures \ref{figChiRG} and  \ref{figNuRG}. The figures for the models RG2 and RG3 are similar to those for RG1 and are not reported here.  The Prandtl number in the Solar radiative zone is at any point between $1 \times 10^{-6}$ (the value near the outer boundary) and $2 \times 10^{-5}$ (core); in the RG1 model it is between $1 \times 10^{-7}$ and $5 \times 10^{-7}$. The noticeable spikes in figure \ref{figChiRG} arise from the details of the opacity function. 

\subsection{Shear required to trigger the GSF instability} \label{sec:numresults}
We wish to determine the minimum shear required to trigger the GSF instability in a rotating star.  In the case at hand, this may be done rather simply by
evaluating the sign of the last term of the dispersion relation by Goldreich \& Schubert, given in equation \eqref{GS3}.  We apply this to the stability problem of the radiative zones of the Sun and of the RG models we have earlier described. At a given location, the result will depend on the structural variables of the star (e.g. $\rho, T, \chi, \nu$), the angular velocity $\Omega$, and the shear. In the $\nu = 0$ case, of course, any shear in the $\hat z$ direction would suffice to induce instability.  In what follows, it is convenient to express the shear in dimensionless form. In cylindrical coordinates, this amounts to calculating $\partial \log \Omega / \partial \log R$ and $\partial \log \Omega / \partial \log z$. However, rotation in spherical shells is often assumed to be a good approximation in the radiative zone of a star, so that only the $r$-derivative $\partial \log \Omega / \partial \log r$ is required.

While the observations from asteroseismology provide estimates of $\Omega$, the shear is much less precisely determined. The data on the radiative zone of the Sun are consistent with uniform rotation, with shear present only in the upper boundary adjacent the convective zone\footnote{In this paper, when we refer to the ``shear'' in the text, we imply that we mean its absolute value. In all cases of interest to us, $\partial \log \Omega / \partial \log r$ is negative.}. A straightforward interpolation of a recent set of helioseismology data gives, at $r = 0.70 $R$_\odot$ \citep{Caleo2015}:
\begin{equation} \label{shearSun}
\frac{\partial \log \Omega}{\partial \log R} = - 0.11 , \qquad \frac{\partial \log \Omega}{\partial \log z} = - 0.24 .
\end{equation}
The data are even less constraining for the RGs. \citet{Deheuvels2014} identified two subgiants for which the observations are significantly better reproduced by a curve with a discontinuity in $\Omega$ at a location near the H-burning shell, rather than by a smooth model. The smooth curve would correspond to a shear (inferred from their figure 10) of order: $\partial \log \Omega / \partial \log r \sim -1$. The shear near the discontinuity would obviously be larger than this, but it is not currently possible to estimate it with accuracy. 

The structural quantities for our models have been interpolated from the \texttt{PROSECCO} models. For all models, we adopt an angular velocity value which is both a good approximation for the radiative zone of the Sun, and a reasonable average value for the radiative zone of a typical RG: $\Omega = 2.7 \times 10^{-6}$ rad s$^{-1}$. At about 20 radial locations in the radiative zone of each star, at low (20\degree), mid (45\degree), and high (70\degree) latitude, we solve equation \eqref{GS0} for a range of values of $k_R, k_z$, for the 4 sign combinations of $k_R$, $k_z$. In this way, we determine the minimum value of $\partial \log \Omega / \partial \log r$ for which there are solutions to the equation $C(\bb k) < 0$. We consider wave vector components in the range:
\begin{equation}  \label{krange}
k_{R}, k_{z}: \ \pm \frac{2 \pi}{10^{-2} R_\odot} \rightarrow \pm \frac{2 \pi}{10^{-14} R_\odot},
\end{equation}
and limit our search to values of the shear in the interval:
\begin{equation} \label{shearlimits}
0.1 < \Big| \frac{\partial \log \Omega}{\partial \log r} \Big| < 10 .
\end{equation}

\begin{figure}
	\centering
	\includegraphics[width=0.475\textwidth, clip=true, trim=0cm 0cm 0cm 0cm]{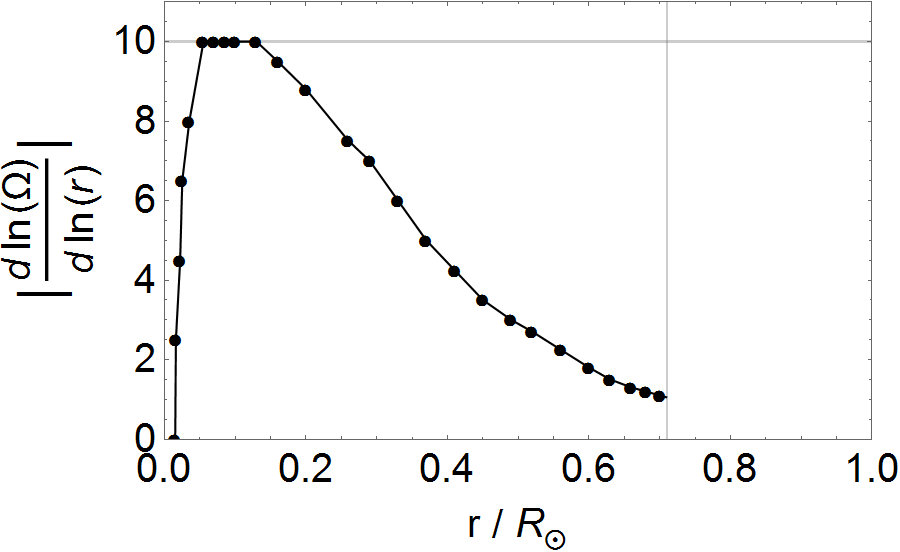}					
	\caption{\label{figShearSun}Minimum shear required for the onset of the GSF instability in the Sun, as a function of the distance from the centre of the star. The vertical grey line shows the boundary between the radiative and convective zone. The thin black line is a linear interpolation of the data points. The calculation is limited to the values of equation \eqref{shearlimits}. The horizontal grey line shows the upper value $d \log \Omega / d (\log r) = 10$.}
\end{figure}

\begin{figure}
	\centering
	\includegraphics[width=0.475\textwidth, clip=true, trim=0cm 0cm 0cm 0cm]{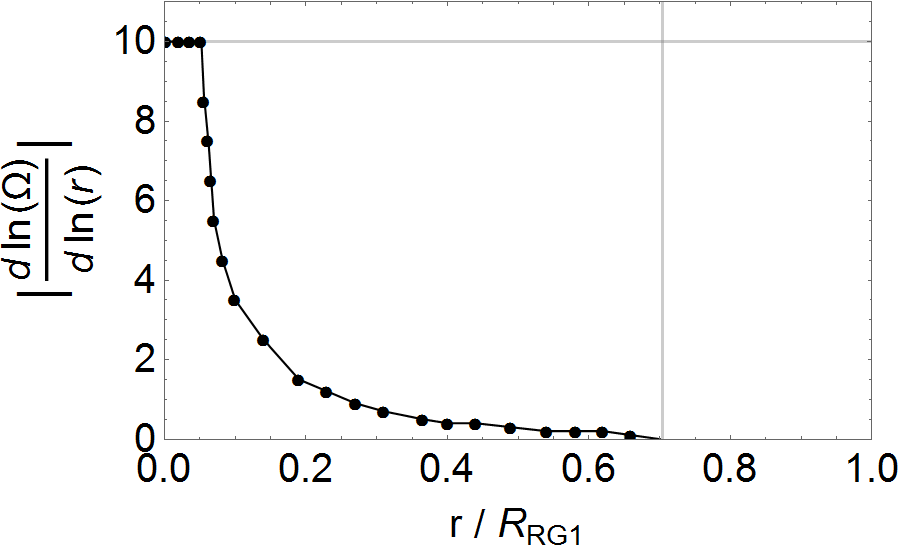}					
	\caption{\label{figShearRG1}Minimum shear required for the onset of the GSF instability in the RG1 model. The vertical grey line shows the boundary between the radiative and convective zone.}
\end{figure}

\begin{figure}
	\centering
	\includegraphics[width=0.475\textwidth, clip=true, trim=0cm 0cm 0cm 0cm]{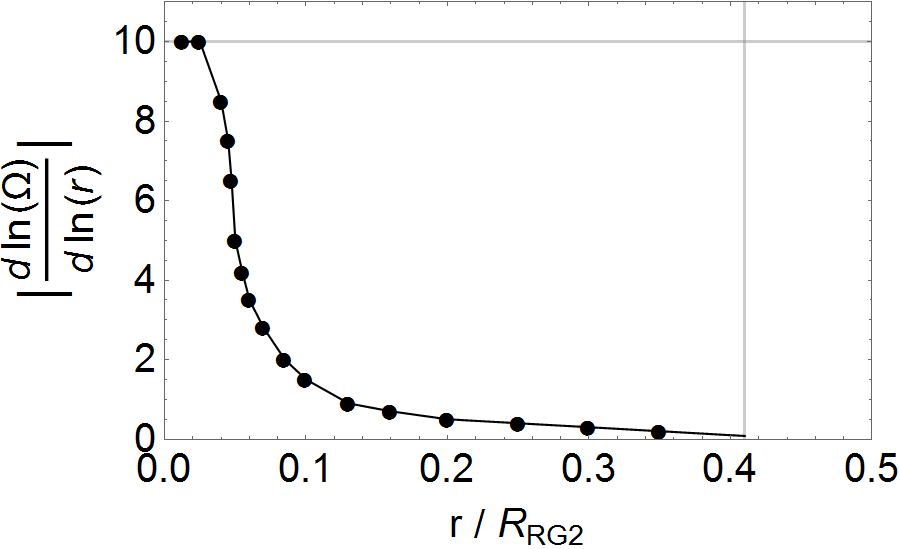}					
	\caption{\label{figShearRG2}Minimum shear required for the onset of the GSF instability in the RG2 model. The vertical grey line shows the boundary between the radiative and convective zone.}
\end{figure}

\begin{figure}
	\centering
	\includegraphics[width=0.475\textwidth, clip=true, trim=0cm 0cm 0cm 0cm]{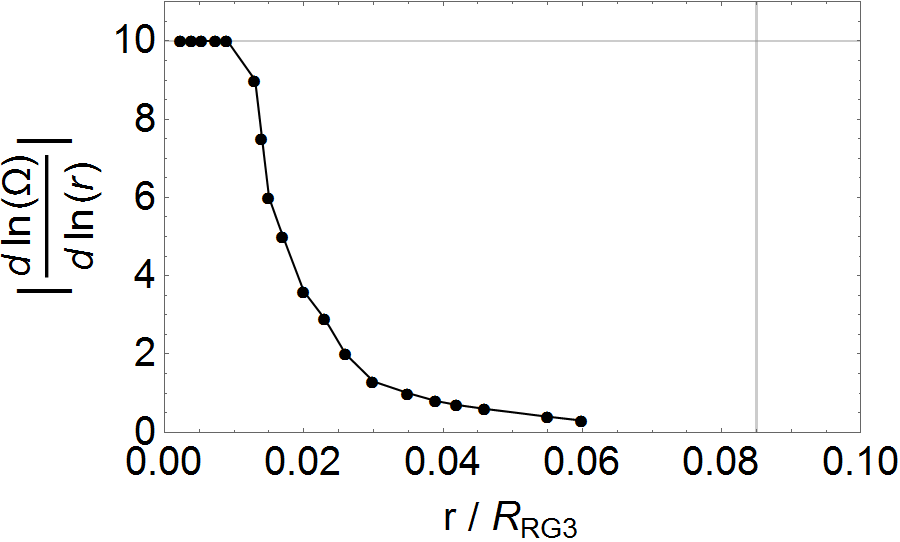}					
	\caption{\label{figShearRG3}Minimum shear required for the onset of the GSF instability in the RG3 model. The vertical grey line shows the boundary between the radiative and convective zone.}
\end{figure}

We show our results in figure \ref{figShearSun} for the Sun and figures \ref{figShearRG1} - \ref{figShearRG3} for the RGs. The main feature implied by these figures is that in all cases the onset of the GSF instability in the deep radiative interior is only possible for very strong shear, while it may occur more easily near the outer edge of the radiative zone. In the Sun, a shear of order unity is required even at the upper boundary of the radiative zone. This is consistent with the already noted result that the GSF instability does not occur in the upper radiative zone, for values of the shear given by equation \eqref{shearSun} (CB16). On the other hand, a small shear $\partial \log \Omega / \partial \log r \sim 0.1$ may be sufficient to induce instability in the upper radiative zone of the RG models, or in the very inner part of the nuclear core of the Sun.


\section{Non-axisymmetric perturbations and introduction of a background magnetic field} \label{sec:appendixGS}
The GSF instability is traditionally studied in the axisymmetric case. The evolution of non-axisymmetric displacements, i.e. displacements with a finite azimuthal component of the wave vector $k_\phi$, is a much more complex problem. 
In this case, the perturbed equations cannot be reduced to a local  dispersion relation. This result is well known to the fluid dynamicists and need not be elaborated upon here. We refer the reader to CB16 for a discussion of this issue and further references, and adopt here the same notation as in that paper. There we have derived the set of non autonomous ordinary differential equations which describe the evolution of the perturbed quantities in the non-axisymmetric case, see equations (51) - (55). We also make use of the same standard, non-rotating background solar model as in that paper \citep{BahcallSerenelliBasu2005}.

The possible importance of non-axisymmetric perturbations is that they can be more unstable in a rotating system than axisymmetric disturbances.   This happens, for example, in rotating convectively unstable systems in which a geostrophic balance eliminates rotational stabilisation, but only for non-axisymmetric modes \citet{BalbusSchaan2012}.
We therefore discuss here an idealised case in which the GSF instability is not fully suppressed by the viscosity, and solve those equations to determine what happens when the assumptions of axisymmetry and non-magnetised background are relaxed, exploring the richness and complexity of the general, triple-diffusive problem. For this purpose, we consider an environment with the same properties of the upper radiative zone of the Sun (including the rotation), but an enhanced radiative thermal diffusion coefficient: $\xi_{\text{rad}} = 10^3 \xi_{\text{rad} \odot}$. For clarity, we note that the coefficient $\xi_{\text{rad}}$ of CB16 is related to the coefficient $\chi$ of the current paper by $\xi_\text{rad} = (\gamma - 1) \chi$.

\subsection{Axisymmetric perturbations}
\begin{figure}
	\centering
	\includegraphics[width=0.5\textwidth, clip=true, trim=0cm 0cm 0cm 0cm]{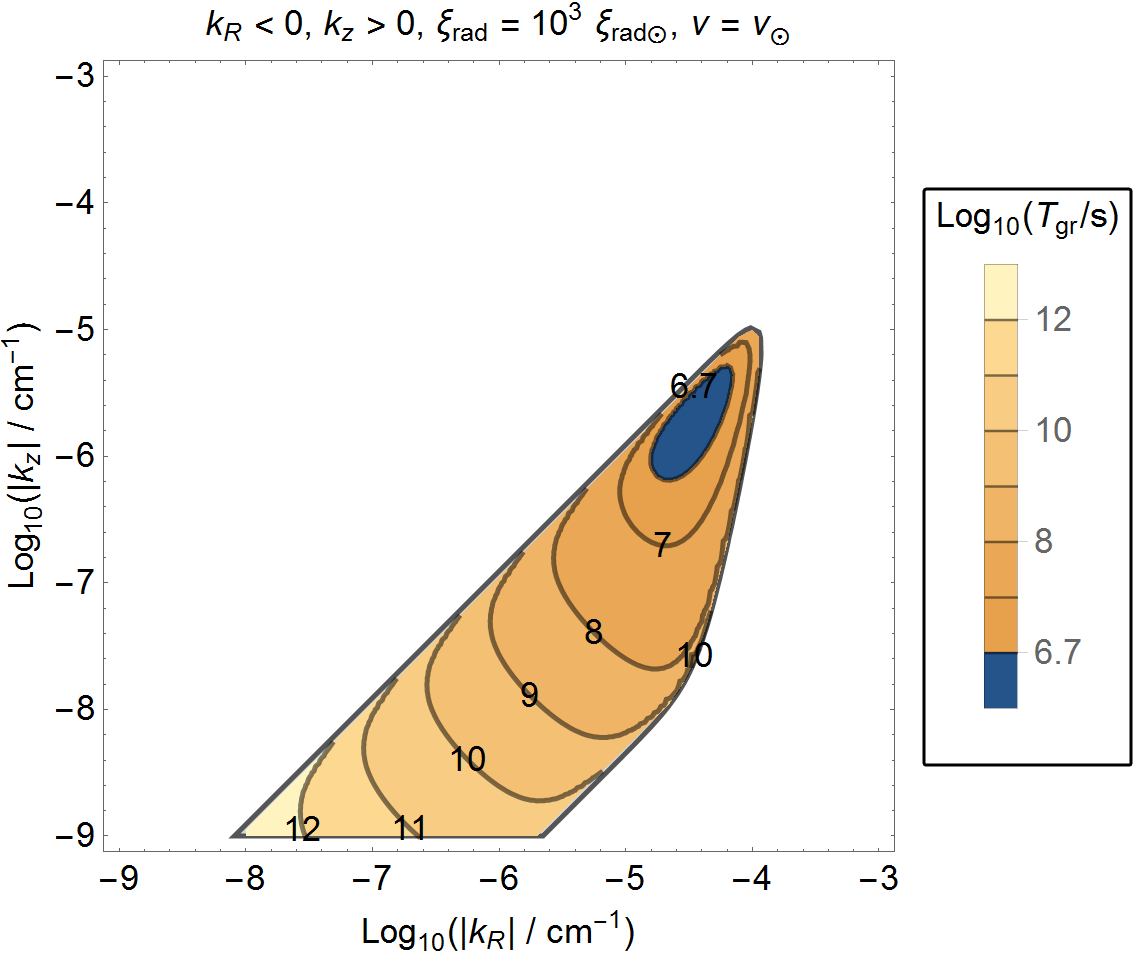}			\caption{\label{figGS3}Region of the $k_R < 0, k_z > 0$ plane where the GSF instability occurs in the Sun and growth time-scale $T_{\text{gr}}$ for the axisymmetric displacements, for $r = 0.7 R_\odot$, $\theta = 45 \degree$, with $\xi_{\text{rad}} = 10^3 \xi_{\text{rad} \odot}$ and $\nu = \nu_\odot$. The numbers next to the iso-contours correspond to the value of $\log_{10}(T_{\text{gr}})$ with $T_{\text{gr}}$ expressed in seconds.}
\end{figure}

The axisymmetric case is treated with the techniques of the present paper. We have solved equation \eqref{GS0} in the upper radiative zone of the Sun at $r = 0.70$ R$_\odot$, $\theta = 45 \degree$, with the modification $\xi_{\text{rad}} = 10^3 \xi_{\text{rad}\odot}$. There are unstable modes only when $k_R$ and $k_z$ have opposite signs, and the results for $k_R > 0, k_z < 0$ are the same as those for $k_R < 0, k_z > 0$. Only modes in a relatively narrow region of the $k_R < 0, k_z > 0$ quadrant are unstable: we show this region and the growth time-scale of the instability in figure \ref{figGS3}. The shortest growth time-scale is found to be given by $\log_{10}(T_{\text{gr}} / s) \approx 6.7$, i.e. $T_{\text{gr}} \approx 5 \times 10^6$ s, with a strong dependence of $T_{\text{gr}}$ on the position in the $k_R - k_z$ plane. 

\subsection{Non-axisymmetric perturbations}
As noted above, the evolution of perturbations with finite $k_\phi$ is not described by a plane wave dispersion relation, but rather determined by solving equations (51) - (55) of CB16. However, a qualitative understanding of their behaviour can be obtained by noting that perturbations with a finite but small $k_\phi$ will still show a predominantly axisymmetric behaviour, albeit with values of $k_R$ and $k_z$ that are not constant in time. The classical dispersion relation by GS67 is therefore still of some use in understanding how these displacements behave, provided that the time dependence of $k_R$ and $k_z$ is correctly included in it.

At large times, all perturbations formally evolve towards a quasi-axisymmetric state with a wave vector of the form (see CB16, equation (16)):
\begin{equation}  \label{eventualkRkz}
k_R (t) \rightarrow - A k_\phi t \frac{\partial \Omega}{\partial R}, \qquad k_z (t) \rightarrow - A k_\phi t \frac{\partial \Omega}{\partial z} ,
\end{equation}
for some value of $A$.  Equivalently, they evolve towards a state with
\begin{equation}  \label{eventualkRkzratio}
\frac{k_R}{k_z} = \frac{\partial \Omega / \partial R}{\partial \Omega / \partial z} .
\end{equation}
We explored the stability of axisymmetric perturbations that adhere to the constraint \eqref{eventualkRkzratio} in the range of amplitudes $|\bb k| = 10^{-9} - 10^{-1} $cm$^{-1}$ and co-latitudes $\theta = 10 \degree - 80 \degree$. We found no unstable modes: \emph{all the non-axisymmetric modes are eventually stable}.

The next natural step of this analysis is to study the transient phase to assess the presence of large initial growths that occur before the perturbation moves out of the unstable region of the $k_R - k_z$ plane. The following empirical argument allows us to estimate a threshold for $|k_\phi|$ for the perturbation to grow by a large factor.  We may visualize the position of the perturbation as a point in figure \ref{figGS3}, which moves over the course of time. The perturbation will have a large growth phase if it remains in an unstable region of the plane for a time that is much longer than the growth time-scale in such region. The size of the unstable region of the plane is bounded by $|\Delta k| \lesssim 10^{-4} $cm$^{-1}$. The position of the perturbation in the region changes with constant wavenumber velocity:
\begin{equation}  \label{kRkzvelocities}
\dot k_R = - k_\phi R \frac{\partial \Omega}{\partial R}, \qquad \dot k_z = - k_\phi R \frac{\partial \Omega}{\partial z} .
\end{equation}
The condition that the perturbation remains in the region of close to maximum growth for a time much longer than the growth time-scale, say $\Delta t = 10 T_{\text{gr}}$, gives the constraint:
\begin{equation} 
|k_\phi| < \frac{\Delta k}{R |\bb \nabla \Omega|} \frac{1}{\Delta t} \lesssim 10^{-6} \text{cm}^{-1} .
\end{equation}
In fact, a detailed exploration of the modes in the unstable region shows that only modes with $|k_\phi| \lesssim 10^{-7}$ cm$^{-1}$ are able to grow by many orders of magnitude. Since the most unstable region of the plane resides at $k_R > 10^{-5}$ cm$^{-1}$, only displacements that are strongly axisymmetric are found to be unstable enough for the GSF instability to have the time required to affect them. 

\begin{figure}
	\centering
	\includegraphics[width=0.5\textwidth, clip=true, trim=0cm 0cm 0cm 0cm]{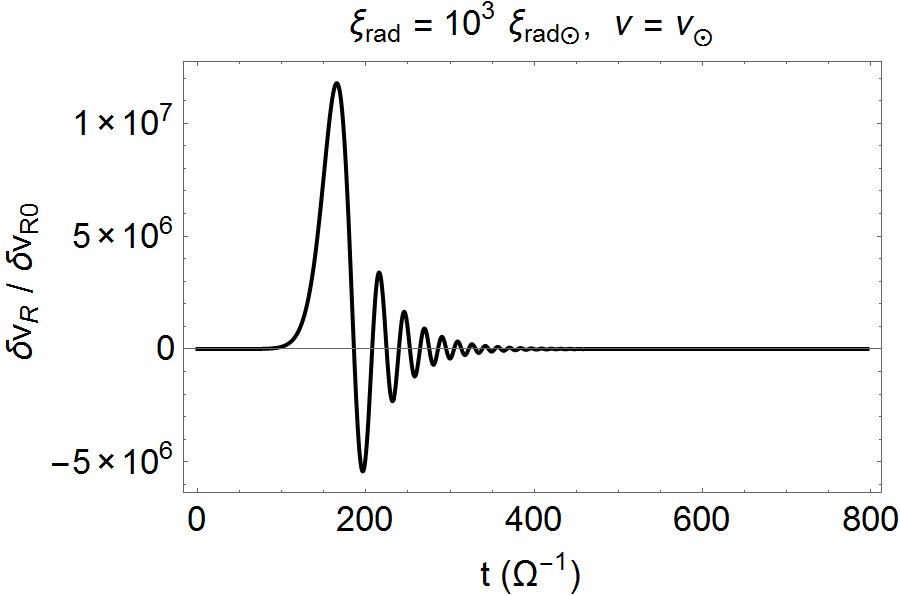}			\caption{\label{figGS4}Evolution of $\delta v_R(t)$ for a perturbation with $k_{R0} = - 10^{-4.5}$ cm$^{-1}$, $k_{\phi} = 10^{-7}$ cm$^{-1}$, and $k_{z0} = 10^{-7}$ cm$^{-1}$. The unit on the temporal axis is expressed in terms of $\Omega^{-1} = 3.77 \times 10^5$ s, corresponding to about 4.4 days.}
\end{figure}

\begin{figure}
	\centering
	\includegraphics[width=0.5\textwidth, clip=true, trim=0cm 0cm 0cm 0cm]{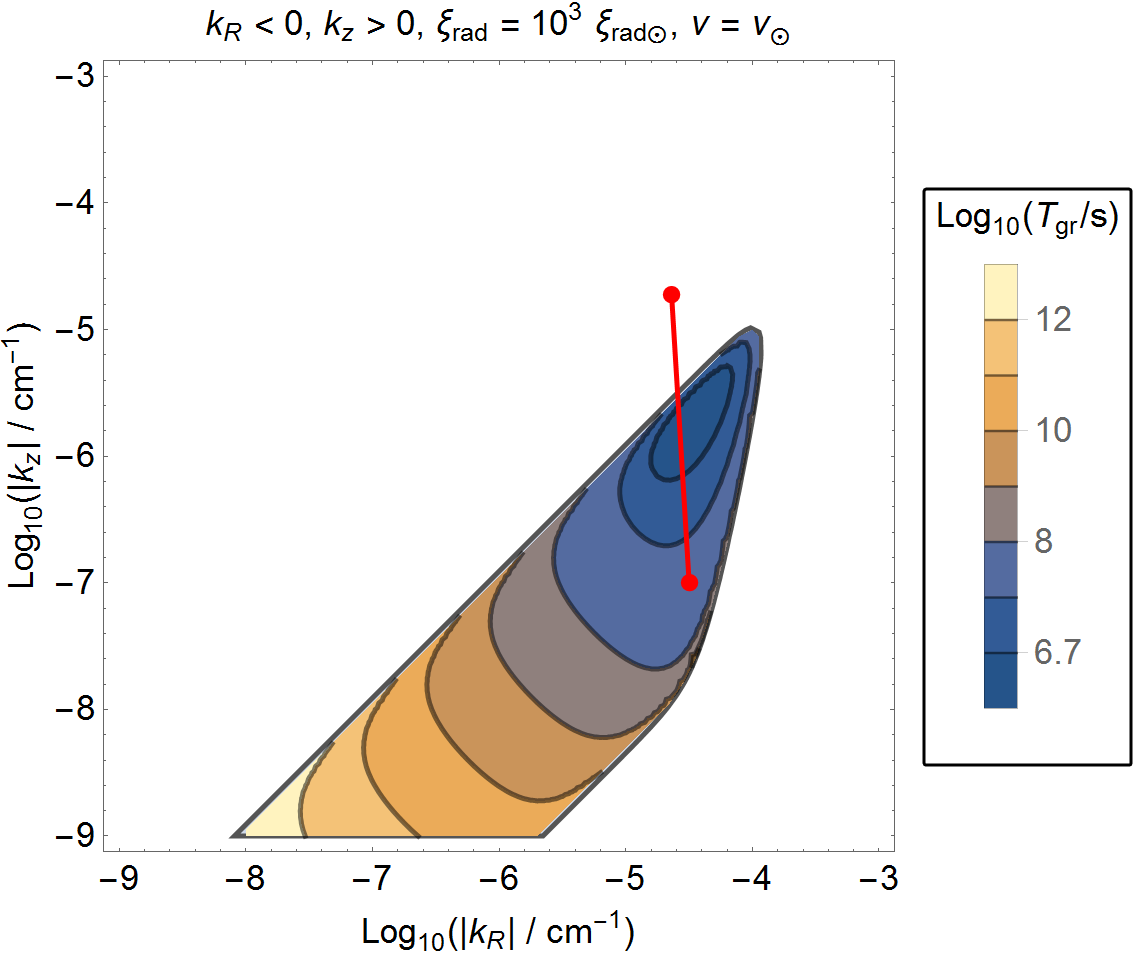}		
	\caption{\label{figGS5}Path tracked by the perturbation of figure \ref{figGS4} in the $k_R - k_z$ plane. The initial position of the perturbation is marked by the red dot in the unstable region of the plane, while the final position is marked by the other red dot. The position of the wave vector does not proceed linearly with time on the segment, due to the logarithmic scale adopted for the axes.}
\end{figure}

\begin{figure}
	\centering
	\includegraphics[width=0.5\textwidth, clip=true, trim=0cm 0cm 0cm 0cm]{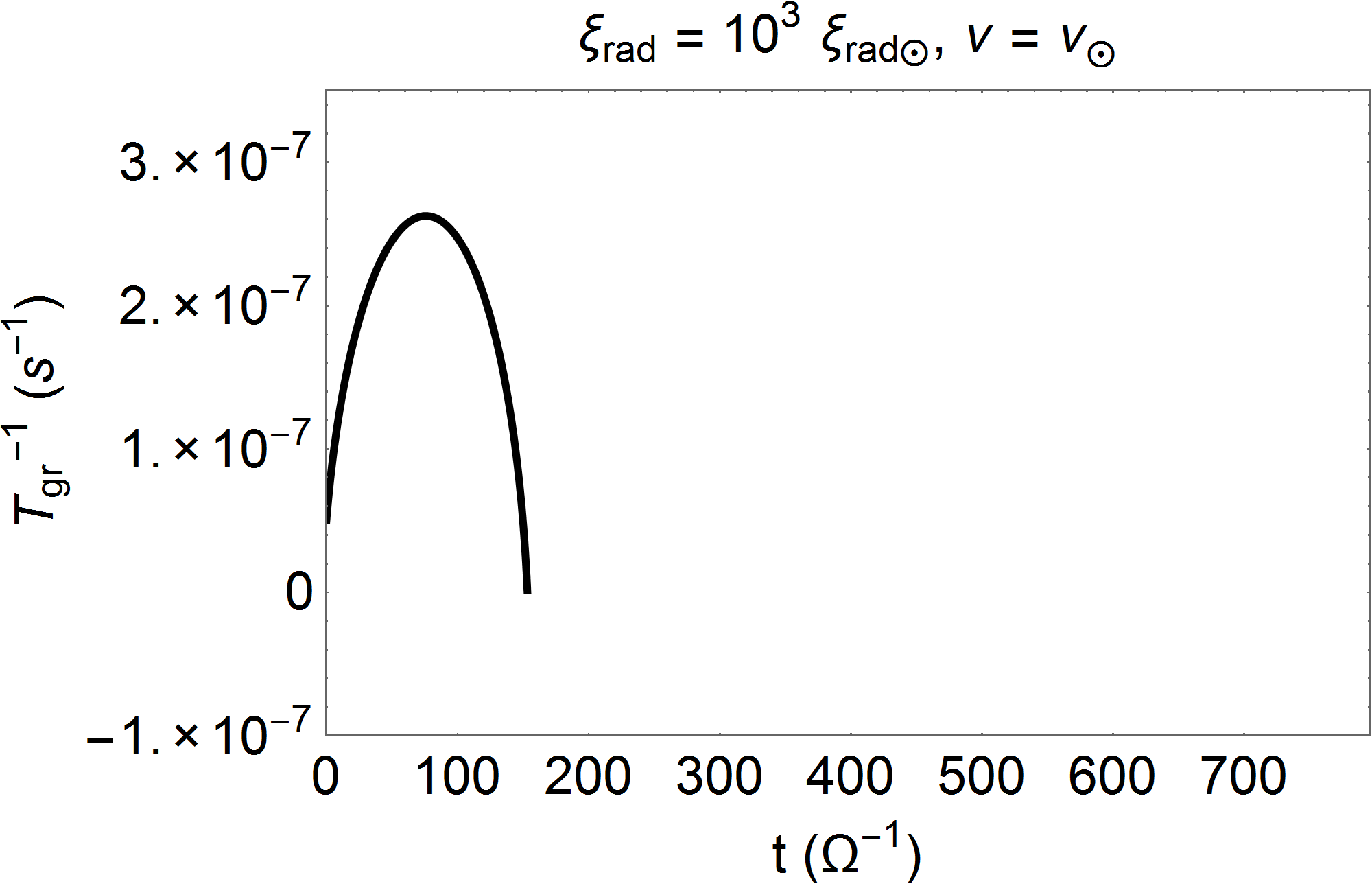}			\caption{\label{figGS6}Growth rate $T_{\text{gr}}^{-1}$ for the displacement of figure \ref{figGS4}, reported for as long as it stays in the unstable region of the $k_R - k_z$ plane. The growth of the perturbation is halted when the growth rate reaches 0. The unit on the temporal axis is as in figure \ref{figGS4}.}
\end{figure}

It is possible to identify perturbations that grow by a large factor before becoming stable. This may be understood on the basis of the axisymmetric theory. We show in figure \ref{figGS4} the evolution of $\delta v_R (t) / \delta v_R(0)$ for a perturbation with initial wave vector components $k_{R0} = - 10^{-4.5}$ cm$^{-1}$, $k_{\phi} = 10^{-7}$ cm$^{-1}$, and $k_{z0} = 10^{-7}$ cm$^{-1}$. As the wave vector of the perturbation changes, it eventually moves out of the unstable region of the $k_R - k_z$ plane. The growth is reversed when the perturbation emerges from the unstable region. We show in figure \ref{figGS5} the path of the perturbation in the $k_R - k_z$ plane. Finally, we show in figure \ref{figGS6} the growth rate $T_{\text{gr}}^{-1}$ for the time interval in which the perturbation is in the unstable region. Non-axisymmetric disturbances are not intrinsically unstable on their own; they show growth only to the extent that the instantaneous poloidal wavenumbers would be unstable in an axisymmetric calculation.

\subsection{The stabilizing effect of $\bb B$} \label{sec:GSmag}
\begin{figure}
	\centering
	\includegraphics[width=0.5\textwidth, clip=true, trim=0cm 0cm 0cm 0cm]{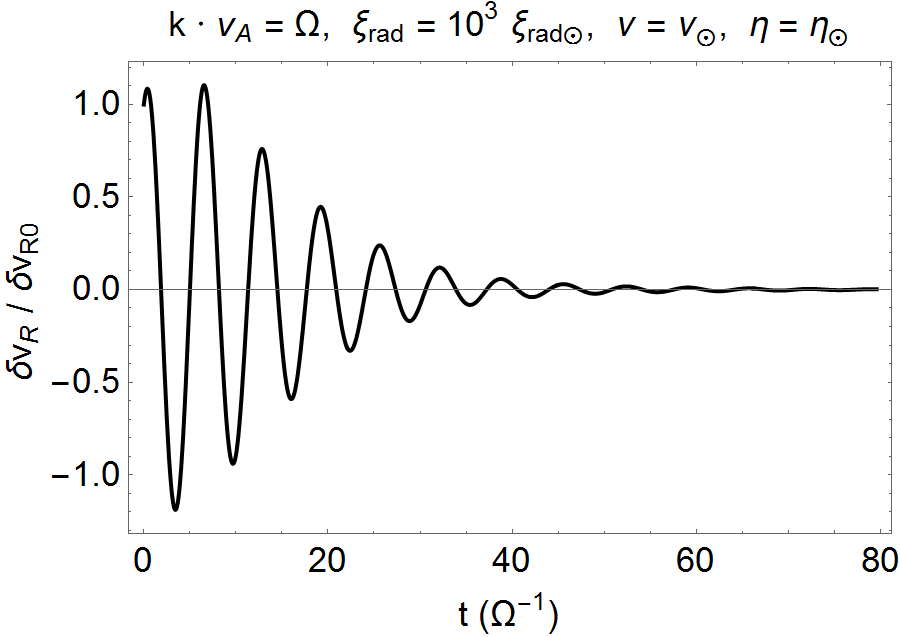}		
	\caption{\label{figGS7}Evolution of $\delta v_R(t)$ for the perturbation with $k_{R0} = - 10^{-4.5}$ cm$^{-1}$, $k_{\phi} = 10^{-7}$ cm$^{-1}$, and $k_{z0} = 10^{-7}$ cm$^{-1}$ in the magnetic case $\bb k \bcdot \bb v_A = \Omega$. The units are as in figure \ref{figGS4}.}
\end{figure}

Finally, we discuss the effect of a finite background magnetic field $\bb B$. $\bb B$ appears in equations (51) - (55) of CB16 only via the constant term $\bb k \bcdot \bb v_A$. It is convenient to compare $\bb k \bcdot \bb v_A$ and $\Omega$. As expected, we found that the behaviour of the perturbation in figures \ref{figGS4} - \ref{figGS6} is unchanged for values of $\bb k \bcdot \bb v_A \ll \Omega$. However, increasing $\bb k \bcdot \bb v_A$ up to values greater than $0.1 \Omega$, we find that the growth is rapidly inhibited, disappearing for $\bb k \bcdot \bb v_A \approx \Omega$. We show in figure \ref{figGS7} the evolution of $\delta v_R (t) / \delta v_R(0)$ for the same perturbation in the case $\bb k \bcdot \bb v_A = \Omega$.

A qualitative understanding of the role of $\bb B$ is as follows. Once the effect of the background magnetic field becomes comparable to that of the rotation, the diffusivity term that most contributes to the damping is the resistivity $\eta$, rather than the viscosity. We have adopted the same value of $\eta$ as in CB16, $\eta \sim 6 \times 10^2$ cm$^2$ s$^{-1}$, so that $\eta$ is at least one order of magnitude larger than $\nu$ in the problem at hand. This makes the damping more efficient in the magnetised case.

Interestingly, the condition $\bb k \bcdot \bb v_A \ge \Omega$ for the wave vector of the perturbation in figures \ref{figGS4} - \ref{figGS6} gives $\bb v_A \gtrsim 0.1$ cm s$^{-1}$, a value many orders of magnitude smaller than the sound speed in the medium and the rotational velocity. The corresponding magnetic field is $|\bb B| \gtrsim 0.2$ G. Even if the radiative diffusion coefficient were much higher than in the Sun, a very small magnetic field would be sufficient to suppress the GSF instability.

\section{Conclusions}
The GSF instability is often considered to be one of the sources of AM transport in stellar interiors. It is thought to play a (modest) role in determining the time evolution the angular velocity $\Omega(r, \theta)$ in the star, and to affect the mixing of chemical elements in the upper radiative zone of stars in the AGB.

This instability is typically incorporated in codes of stellar evolution in a diffusion-like approximation assuming a non-viscous background. We have shown here that this approximation is not supported by a detailed analysis of the linear stability problem: when realistic values of the kinematic viscosity are accounted for, the GSF instability is suppressed in the bulk of the radiative zone of both the Sun and RGs at various evolutionary stages.

Finally, in a specific case of an environment which would normally be GSF-unstable, we have investigated the effect of a small deviation from axisymmetry and the presence of a small background magnetic field. Both these effects appear to have a stabilising influence.

\bibliographystyle{mn2e}
\bibliography{References}
\bibdata{References}
%\bibdata{PmInstability}
%bibstyle{mn2e}

\label{lastpage}

\end{document}